\documentclass[12pt,aps,final,showpacs,centertags]{revtex4}

\begin{document}

\title{Strength fragmentation of Gamow-Teller transitions and delayed
neutron emission of atomic nuclei}

\author{
\firstname{A. P.}~\surname{Severyukhin}$^{1),2)}$}

\affiliation{
$^{1)}$ {\rm Joint Institute for Nuclear Research, 141980 Dubna, Moscow region, Russia} \\
$^{2)}$ {\rm Dubna State University, 141982 Dubna, Moscow region, Russia}\\
$^{*}$  {\rm E-mail: }{\bf sever@theor.jinr.ru}}

\begin{abstract}
{\scriptsize Starting from a Skyrme interaction with tensor terms,
the $\beta$-decay rates of $^{52}$Ca have been studied within a microscopic
model including the $2p-2h$ configuration effects. We observe a redistribution
of the strength of Gamow-Teller transitions due to the $2p-2h$ fragmentation.
Taking into account this effect results in a satisfactory description
of the neutron emission probability of the $\beta$-decay in $^{52}$Ca.}
\end{abstract}

\pacs{21.60.Jz, 23.40.-s}
\maketitle


The multi-neutron emission is basically a multistep process
consisting of (a) the $\beta$-decay of the parent nucleus (N, Z)
which results in feeding the excited states of the daughter
nucleus (N - 1, Z + 1) followed by the (b) $\gamma$-deexcitation to
the ground state or (c) multi-neutron emissions to the ground
state of the final nucleus (N - 1 - X, Z + 1), see e.g.,
Ref.~\cite{b05}. Predictions of the multi-neutron emission are
needed for the analysis of radioactive beam experiments and
for modeling of astrophysical r-process. Recent experiments
gave an evidence for strong shell effects in exotic calcium
isotopes~\cite{w13,s13}. For this reason, the $\beta$-decay
properties of neutron-rich isotope $^{52}$Ca provides valuable
information~\cite{h85}, with important tests of theoretical calculations.

%
\begin{table}[]
\caption[]{$\log ft$ values and excitation energies of the low-lying $1^+$ states populated
in the $\beta$-decay of $^{52}$Ca. Results of the calculations without the two-phonon effects (QRPA)
and with the two-phonon effects (2PH) are shown. Experimental data are taken from Ref.~\cite{h85}.}
\begin{tabular}{ccccccc}
\noalign{\smallskip}\hline
$\lambda_i^{\pi}=1_i^+$&\multicolumn{3}{c}{$E_x$ (MeV)}&\multicolumn{3}{c}{$\log ft$}    \\
                       & Expt. & QRPA & 2PH            & Expt.     & QRPA & 2PH          \\
\hline\noalign{\smallskip}
$1_1^+$                &1.64   & 1.5  & 1.3            &4.2$\pm$0.1& 4.3  &  4.3         \\
$1_2^+$                &2.75   &      & 3.9            &4.5$\pm$0.2&      &  6.4         \\
$1_3^+$                &3.46   &      & 4.2            &5.3$\pm$0.5&      &  9.2         \\
$1_4^+$                &4.27   & 5.0  & 4.9            &4.0$\pm$0.5& 3.2  &  3.3         \\
\noalign{\smallskip}\hline
\end{tabular}
\end{table}
One of the successful tools for studying charge-exchange nuclear
modes is the quasiparticle random phase approximation (QRPA) with
the self-consistent mean-field derived from a Skyrme
energy-density functional (EDF) since these QRPA calculations
enable one to describe the properties of the parent ground state
and Gamow-Teller (GT) transitions using the same EDF. Making use of
the finite rank separable approximation (FRSA) \cite{gsv98} for the
residual interaction, the approach has been  generalized for the
coupling between one- and two-phonon components of the wave
functions~\cite{svg04}. The FRSA in the cases of the charge-exchange
excitations and the $\beta$-decay was already introduced in
Refs.~\cite{svg12,ss13} and in Ref.~\cite{svbag14,e15}, respectively.
In the case of the $\beta$ decay of $^{52}$Ca, we use
the EDF T45 which takes into account the tensor force added with
refitting the parameters of the central interaction~\cite{TIJ}. The
pairing correlations are generated by a zero-range volume force
with a strength of -315 MeVfm$^{3}$ and a smooth cut-off at 10 MeV
above the Fermi energies~\cite{svbag14}. This value of the pairing
strength has been fitted to reproduce the experimental neutron
pairing energy of $^{52}$Ca obtained from binding energies of
neighbouring Ca isotopes.

Taking into account the basic ideas of the quasiparticle-phonon model
(QPM)~\cite{solo,ks84}, the Hamiltonian is then diagonalized in a
space spanned by states composed of one and two QRPA phonons~\cite{svbag14},
\begin{eqnarray}
\Psi _\nu (J M) = \left(\sum_iR_i(J \nu )Q_{J M i}^{+}+
\sum_{\lambda _1i_1\lambda _2i_2}P_{\lambda _2i_2}^{\lambda
_1i_1}( J \nu )\left[ Q_{\lambda _1\mu _1i_1}^{+}\bar{Q}_{\lambda
_2\mu _2i_2}^{+}\right] _{J M }\right)|0\rangle, \label{wf}
\end{eqnarray}
where $Q_{\lambda \mu i}^{+}\mid0\rangle$ are  the wave functions of the one-phonon states
of the daughter nucleus (N - 1, Z + 1); $\bar{Q}_{\lambda\mu i}^{+} |0\rangle$ is the
one-phonon excitation of the parent nucleus (N, Z). We use only the two-phonon
configurations $[1^{+}_{i}\otimes 2^{+}_{i'}]_{QRPA}$.
In the allowed GT approximation, the $\beta^{-}$-decay rate is expressed
by summing the probabilities (in units of $G_{A}^{2}/4\pi$) of the energetically
allowed transitions  ($E_{k}^{\mathrm{GT}}\leq Q_{\beta}$) weighted with the integrated
Fermi function
\begin{eqnarray}
T_{1/2}^{-1}=D^{-1}\left(\frac{G_{A}}{G_{V}}\right)^{2}
\sum\limits_{k}f_{0}(Z+1,A,E_{k}^{\mathrm{GT}})B(GT)_{k},
\end{eqnarray}
\begin{equation}
 E_{k}^{\mathrm{GT}}=Q_{\beta}-E_{1^+_k},
\end{equation}
where $G_A/G_V$=1.25 and $D$=6147 s. $E_{1_k^+}$ denotes
the excitation energy of the daughter nucleus. As proposed in
Ref.~\cite{ebnds99}, this energy can be estimated by the following
expression:
\begin{equation}
E_{1^{+}_{k}}\approx E_{k}-E_{\textrm{2QP},\textrm{lowest}}.
\end{equation}
$E_{k}$ are the eigenvalues of the wave functions~(\ref{wf})
and $E_{\textrm{2QP},\textrm{lowest}}$ corresponds
the lowest two-quasiparticle energy. The difference in the
characteristic time scales of the $\beta$ decay and subsequent
particle emission processes justifies an assumption of their
statistical independence (see Ref.~\cite{b05} for more details).
The $P_{n}$ probability of the delayed neutron emission is defined
as the ratio of the integral $\beta$-strength to the excited states
above the neutron separation energy of the daughter nucleus.

The spectrum of four low-energy $1^+$ states of $^{52}$Sc is shown
in Table 1. The structure peculiarities are reflected in the $\log ft$ values.
We find that the dominant contribution in the wave function of the first
(fourth) $1^+$ state comes from the configuration $\{\pi1f_{7/2}\nu1f_{5/2}\}$
($\{\pi1f_{7/2}\nu1f_{7/2}\}$). The inclusion of the four-quasiparticle configurations
$\{\pi1f_{7/2}\nu1f_{5/2} \nu2p_{3/2}\nu2p_{1/2}\}$ and
$\{\pi1f_{7/2}\nu1f_{5/2} \nu2p_{3/2}\nu2p_{3/2}\}$
plays the key role in our calculations of the
states $1_{2}^+$ and $1_{3}^+$, respectively. The inclusion of
the two-phonon configurations results in the $P_{n}$ value of 5\%,
and the quantitative agreement with the experimental data~\cite{h85}
is satisfactory. Note that this value is almost three times less
than that within the one-phonon approximation.

In summary, by starting from the Skyrme mean-field calculations
the GT strength in the $Q_{\beta}$-window has been studied within
the model including the $2p-2h$ fragmentation. We analyze this effect
on the $\beta$-transition rates in the case of $^{52}$Ca. Including the
$2p-2h$ configurations leads to qualitative agreement with existence
of four low-energy $1^+$ states of $^{52}$Sc. As a result, the probability
of the delayed neutron emission is decreased.

I would like to thank I.N. Borzov, Yu.E. Penionzhkevich,
and D. Verney for fruitful collaboration,
N.N. Arsenyev and E.O. Sushenok for help.
This work is partly supported by CNRS-RFBR Agreement No.~16-52-150003, the
IN2P3-JINR agreement, and RFBR Grant
No.~16-02-00228.

\end{document}